\documentclass[twocolumn,showpacs,aps]{revtex4}
\usepackage{graphicx}% Include figure files

\begin{document}
%\draft %prints PACS numbers in

\title{Nonlinear stability of oscillatory wave fronts in chains
of coupled  oscillators}
\author{A. Carpio\cite{carpio:email} }
\affiliation{Departamento de Matem\'{a}tica Aplicada, 
Universidad Complutense de Madrid, 28040 Madrid, Spain}
\date{ \today   }

\begin{abstract}
We present a stability theory for kink propagation in chains of
coupled oscillators and a new algorithm for the numerical study of
kink dynamics. The numerical solutions are computed using an equivalent 
integral equation instead of a system of differential equations. This 
avoids uncertainty about the impact of artificial boundary conditions 
and discretization in time. Stability results also follow from
the integral version. Stable kinks have a monotone leading edge and 
move with a velocity larger than a critical value which depends on
the damping strength.
\end{abstract}

\maketitle

%\begin{multicols}{2}
%\narrowtext

\section{Introduction}

The dynamics of waves in chains of coupled oscillators is the key 
to understanding the motion of defects in many physical and biological 
problems: motion of dislocations \cite{fk,nab67} or cracks \cite{sle81} 
in crystalline materials, atoms adsorbed on a periodic substrate 
\cite{cha95}, motion of electric field domains and domain walls in 
semiconductor superlattices \cite{bon02}, pulse propagation through 
myelinated nerves \cite{sleeman} or cardiac cells \cite{kee98}... 
A peculiar feature of these spatially discrete systems is that wave
fronts and pulses get pinned for entire intervals of a control
parameter such as an external force. Typically, wave fronts do not
move unless the external force surpasses a control value. Such is
the case with the static and dynamic Peierls stresses in dislocation
dynamics  \cite{hob65,nab67} or the dynamic and static friction 
coefficients \cite{ger01} in continuum mechanics. Pinning and motion 
of wave fronts also explain the relocation of static electric field 
domains and the self-oscillations of the current in semiconductor 
superlattices \cite{bon02}.

Wave front motion in systems of nonlinear oscillators modeling these
phenomena are easier to analyze in the overdamped case, and less so if
inertia is important. In the presence of inertia, the wrong choice of
boundary conditions or the numerical method may suppress important 
solutions of the original system or yield spurious oscillations.
Thus two problems that are important in all spatially discrete systems 
acquire even more importance: how do we find wave fronts? and, what are 
their stability properties?

We have solved the first problem in a recent work \cite{wavy} by
choosing a damped system of oscillators with a piecewise linear 
source term, see also  \cite{atk65,sle81,truski}. 
Our results show explicitly the existence of kinks with
oscillatory profiles for systems with little or no damping.
In the latter case, these wave fronts have at least one tail
with non decaying oscillations that extend to infinity. 
Depending on the control parameter, branches of oscillatory wave 
fronts may exist, coexisting for entire intervals of the external
force and even coexisting with pinned wave front solutions. 
These facts, long-lived oscillatory profiles and coexistence of wave 
front branches, highlight the importance of ascertaining the 
stability properties of these solutions. 
This is not easy and not many results are known.

To be precise, let us consider oscillator chain:
\begin{eqnarray}
m u_n'' \!+\! \alpha u_n' \!=\! K(u_{n+1}-2u_n +u_{n-1}) \!-\!V'(u_n) \!+\! W.
\label{e01}  
\end{eqnarray}
We nondimensionalize the model by choosing the time scale $\sqrt{{m a^2\over v_m}}$,
where $a$ is the interatomic distance, $m$ is the mass and $v_m$ the strength
of the on-site potential. For a piecewise parabolic potential, the nondimensional
equation is:
\begin{eqnarray}
 u_n'' + \gamma u_n' = D(u_{n+1}-2u_n +u_{n-1}) - g(u_n) + F,
\label{e1} \\
g(s)= \left\{\begin{array}{ll}
s  \quad  s < {1\over 2} \\
s - 1 \quad s \geq {1\over 2}.
\end{array}\right. \label{e2}
\end{eqnarray}
Here $\gamma$ and $D$ are the ratios between friction and inertial forces, and
between the strengths of the harmonic and on-site potentials, respectively. $F=
Wa/v_m$. Atkinson and Cabrera \cite{atk65} conjectured that only two  branches 
of kinks are stable for (\ref{e1})-(\ref{e2}):
\begin{itemize}
\item  a branch of  static kinks for values of the control 
parameter $|F|$ below an static threshold $F_{cs}(D),$ 
\item  a branch of traveling kinks for $|F|$ above a
dynamic threshold $F_{cd}(\gamma,D)\leq F_{cs}(D)$, with speeds 
$c$ larger than a minimum speed $c_{cd}(\gamma,D)$. This family has 
a distinctive feature compared to eventual slower waves \cite{wavy}:
the leading edge of the kink is monotone whereas the trailing edge
may develop oscillations.
\end{itemize}
The values $F_{cs}$ and $F_{cd}$ correspond to the static and dynamic 
Peierls stresses of the literature on dislocations \cite{nab67}.
In the overdamped limit $\gamma \rightarrow \infty$, $F_{cs}=F_{cd}$ 
and stable wave fronts can be found with arbitrarily small speeds 
\cite{fat98}.

\begin{figure}
\begin{center}
\includegraphics[width=8cm]{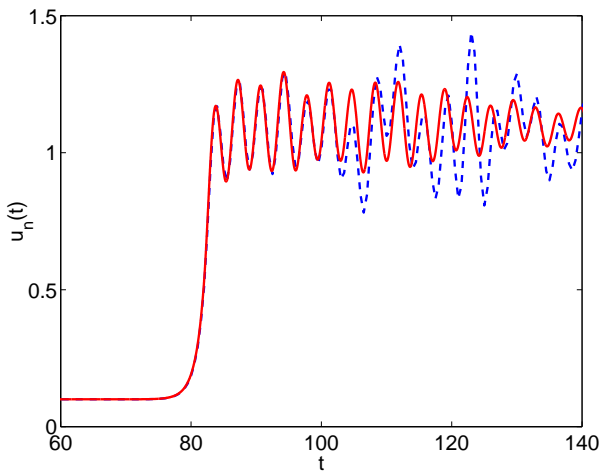}% exi7.eps
\caption{
Trajectory $u_n(t)$ computed by solving a truncated system
of differential equations (dashed) and by integral
expressions (solid) for $\gamma=0.02$, $D=4$, $F=0.1$, $n=-70$. 
}
\label{figura0}
\end{center}
\end{figure}

In a previous paper \cite{wavy}, we checked numerically the validity
of Atkinson and Cabrera's conjecture. This is a delicate affair and 
further analytical work is clearly desirable. In fact, most numerical 
studies of kink propagation truncate the infinite chain 
to a finite chain, fix some boundary conditions and then use a 
Runge-Kutta solver (or variants) to investigate the dynamics of kink-like 
initial configurations. For instance, Peyrard and Kruskal \cite{pey84}
applied this procedure to study kinks in the conservative Frenkel-Kontorova 
model, including friction near the ends of the truncated chain
in an attempt to avoid reflections.
On the other hand, our analytical work \cite{wavy} shows that traveling 
kinks  oscillate with almost uniform amplitude even for small damping. 
Then, artificial boundary conditions and time discretization may greatly 
distort their shape and dynamics. In fact, using Runge-Kutta methods to 
solve equation (\ref{e1}) with constant boundary conditions generates
reflections at the boundary, as shown in Figure \ref{figura0},
after a waiting time depending  on the size of the lattice.
Such oscillations end up distorting the right tail and may completely
alter the shape of the kink giving rise to a complex oscillatory 
pattern.

A good way to avoid the spurious effects of inappropiate boundary 
conditions  is to recast  (\ref{e1}) as an integral equation. 
Integral reformulations provide an analytical expression for the 
solutions of (\ref{e1}) which we use to develop new numerical 
algorithms. Spurious pinning and spurious oscillations are suppressed. 
The introduction of these numerical methods based on integral 
reformulations of (\ref{e1}) is one of our main results.

The main analytical results of this paper concern the nonlinear
stability of stationary and traveling wave fronts in chains of
oscillators. 
Besides leading to good numerical methods, we  have also used the
integral equation formulation to investigate the nonlinear stability
of wave front patterns.  We provide a criterion to decide  whether
certain kink-like initial configurations evolve into stable wave 
front patterns. 
In discrete overdamped models the nonlinear stability of traveling 
wave fronts follows from comparison principles. This strategy was 
applied to the study of domain walls in discrete drift-diffusion models 
for semiconductor superlattices in \cite{car00}.

Common belief is that comparison principles do not hold in models
with inertia. This belief is wrong.
How can we asses the stability of traveling wave fronts 
in such models? For large damping, we can directly compare solutions of 
(\ref{e1}) using its equivalent formulation as an integral equation 
thanks to the positivity of the Green functions. As the damping decreases,  
we can ignore the oscillatory tails of the fronts and compare the monotone 
leading edges of the solutions, which drive their motion. The process of 
comparing solutions is technically more complex than in the overdamped 
case because the Green functions change sign,
and the fronts have oscillatory wakes. Summarizing, there are two key 
ingredients for stability. First, the leading edges of the fronts have to 
be monotone.
Second, the Green functions of the linear problem must be positive 
for an initial time interval, of duration comparable to the time the
front needs to advance one position. This restricts the possible values of
the propagation speed for small damping: only fast kinks
are shown to be stable.   
Our methods are quite general and can be extended to 
Frenkel-Kontorova models with smooth sources \cite{preprint} at the 
cost of  technical complications. 

The paper is organized as follows. In section 2 we introduce a
numerical algorithm and discuss the stability of static kinks.
The stability theory for traveling kinks is presented in Section 3. 
In section 4 we discuss the role of oscillating Green functions
in the appearance of static and dynamic thresholds due to
coexistence of stable static and traveling waves. In section 5
we briefly comment on extensions to oscillator chains with smooth 
cubic sources. Section 6 contains our conclusions. 
Basic details on the pertinent Green functions are recalled in
Appendices A and B. Proofs of our main stability results can be 
found in Appendices C and D.

\section{Static kinks}
\label{sec:static}

\begin{figure}
\begin{center}
\includegraphics[width=8cm]{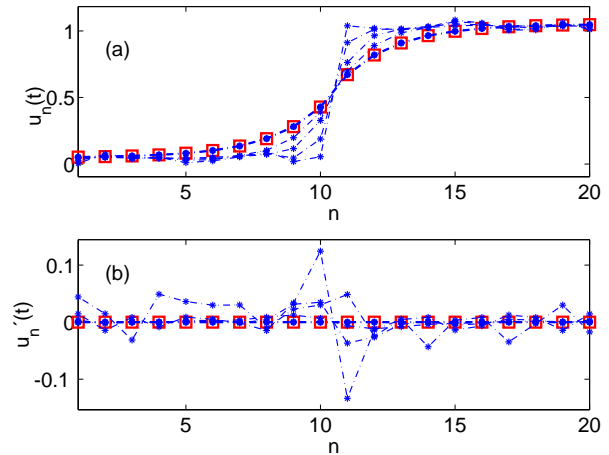}% exi1.eps
\caption{
Convergence to a static kink $s_n$ when $D=4$, $\gamma=10$
and $F=0.05$: (a) asterisks $u_n(t)$, squares 
$s_n$, (b) asterisks $u_n'(t)$, squares $s_n'=0$. }
\label{figura1}
\end{center}
\end{figure}

The stationary wave fronts $s_n$ for (\ref{e1}) increase from $s_{-\infty}=F$ 
to $s_{\infty}=1+F$ and solve the second order difference equation:
\begin{eqnarray}
D(s_{n+1}-2s_n+s_{n-1}) -s_n + H \big(s_n-{1\over 2}\big) +F =0
\label{st0}
\end{eqnarray}
in which $H(x)$ is the Heaviside unit step function. These fronts
are translation invariant.  We fix their position  by setting $s_0 
< {1\over 2} < s_1$. Then, $s_n=F+ a r^n$ for $n\leq 0$ and $s_n=1+F-b r^{-n}$
for $n \geq 1$, where $r ={ 2D+1 +\sqrt{4D+1} \over 2 D}$. Inserting
these formulas in equation (\ref{st0}) for $n=0$ and $n=1$, we find
$a$ and $b$.
Our construction of the stationary fronts $s_n$ is consistent with the
restriction $s_0 < {1\over 2} < s_1$ when $|F|\leq F_{cs}(D)$. Figure 
\ref{figura1} (a) shows a static wave front for $D=4$ and $F=0.05$. 
As $D$ grows, the number of points in the transition layer between the 
constants increases.

\subsection{Stability}

A stationary wave front $s_n$ is stable for the dynamics (\ref{e1}) when 
chains initially close to $s_n$ remain near $s_n$ for all $t>0$, 
as shown in Figure \ref{figura1}. The initial states chosen in this
figure are $u_n^0=F+ \delta^0_n$ when $n\leq 0$, $u_n^0=1+ F+ \delta_n$
when $n\geq 1$ and $u_n^1= \delta_n^1$. Both $\delta_n^1$ and $\delta_n^0$
are small random perturbations.

To find the stable profiles, we proceed as follows. 
Let $u_n^0$ and $u_n^1$ be the initial position and velocity 
of the chain. In terms of Green functions calculated in Appendix A,
$u_n(t)$ is given by (\ref{ap8})  with $f_k(t)=F + H\big(u_k(t)-{1\over 2}\big):$
\begin{eqnarray}\begin{array}{r}
u_n(t)=\sum_k G_{nk}^0(t) u_k^1 + \sum_k G_{nk}^1(t) u_k^0  \\ 
+ \int_{0}^{t}\sum_{k} G_{nk}^0(t-z)  H\big(u_k(z)-{1\over 2}
\big) dz\\ +  F  \int_{0}^{t}\sum_{k} G_{nk}^0(t-z)  dz.
\end{array}\label{st7}
\end{eqnarray}
If initially $u_k^0<{1\over 2}$ for $k\leq 0$, $u_k^0>{1\over 2}$ for $k\geq 1$:
\begin{eqnarray}\begin{array}{r}
u_n(t)=\sum_k \Big[G_{nk}^0(t) u_k^1 + G_{nk}^1(t) u_k^0\Big] +\\ 
 \int_{0}^{t}\sum_{k>0} G_{nk}^0(t-z)  dz
+ F  \int_{0}^{t}\sum_{k} G_{nk}^0(t-z)  dz
\end{array}\label{st1}
\end{eqnarray}
as long as $u_k(t)<{1\over 2}$ when $k\leq 0$, $u_k(t)>{1\over 2}$ when $k\geq 1$.
For $|F|<F_{cs}(D)$, the static wave front $s_n$ with $s_0<{1\over 2}<s_1$
is a solution of equation (\ref{st7}) that satisfies:
\begin{eqnarray}\begin{array}{r}
s_n =\sum_k  G_{nk}^1(t) s_k 
+  \int_{0}^{t}\sum_{k>0} G_{nk}^0(t-z)  dz   \\
+  F  \int_{0}^{t}\sum_{k} G_{nk}^0(t-z)  dz
\end{array}\label{st2}
\end{eqnarray}
for all $t>0$. Subtracting (\ref{st2}) from (\ref{st1}), we obtain:
\begin{eqnarray}
u_n(t)-s_n=\sum_k \Big[G_{nk}^0(t) u_k^1 + G_{nk}^1(t) (u_k^0-s_k)\Big]
\label{st3}
\end{eqnarray}
This expression holds for $t>0$ provided $u_n(t)-{1\over 2}$ does not change sign 
for any $n$ and $t>0$. For which profiles $u_n(t)$ is this true? Let us 
select the initial state of the chain in the set:
\begin{eqnarray}\begin{array}{r}
\sum_{-\infty}^{\infty} |u_n^0-s_n| < M, \quad
\sum_{-\infty}^{\infty} |u_n^1| < M \\
M< R\; {\rm Min }(1,{1\over C_0+C_1}) 
\end{array} \label{st5}
\end{eqnarray}
with $R={\rm Min }(|s_0-{1\over 2}|,s_1-{1\over 2})$ and $C_0,C_1$ to be defined
below. For $\gamma>0$, we show in Appendix B that  $|G_{nk}^0|\leq C_0 
e^{-\eta t}$,  $|G_{nk}^1|\leq C_1 e^{-\eta t}$ with $\gamma>0$.
This  boundedness property of the Green functions and (\ref{st5}) yield:
\begin{eqnarray}
|u_n(t)-s_n| \leq (C_0+C_1) e^{-\eta t} M \label{st6}
\end{eqnarray}
Then,  $|u_n(t)-s_n|<R$ and $u_n(t)-{1\over 2}$ cannot change 
sign for any $t>0$. Moreover, $u_n(t) \rightarrow s_n$ as $t \rightarrow
0$. 

In summary, the static kinks are exponentially and asymptotically stable
in the damped case. Their basin of attraction includes all initial 
configurations $u_n^0$ and $u_n^1$ selected according to (\ref{st5}).
In the conservative case, the static kinks are merely stable, but not
asymptotically stable, because the previous argument with $\gamma=0$, 
$\eta=0$, $C_0=C_1=1$ only yields $|u_n(t)-s_n| \leq 
2 M$ for all times.

In the continuum limit $D \rightarrow \infty$, the number of points
in the transition layer between constants increases and the distance
between points decreases. Then, $s_0$ and $s_1$ tend to ${1\over 2}$ and 
the set of  states (\ref{st5}) attracted by $s_n$ shrinks as $D$ grows. 
It becomes more likely that  initial kinks in the chain propagate for 
a while and finally become pinned at some shifted static kink 
$v_n=s_{n+l}$, $v_{-l}<{1\over 2} < v_{-l+1}$.

\subsection{Numerical algorithm}

Formula (\ref{st7}) can be used to compute numerically the
dynamics of the chain. However, the computational cost is
high, due to the integral terms and the Green functions.
In this section, we exploit the static front solutions $s_n$ to
reduce the cost and derive new formulae for
$u_n(t)$ which clarify the dynamics
of the chain.

\begin{figure}
\begin{center}
\includegraphics[width=8cm]{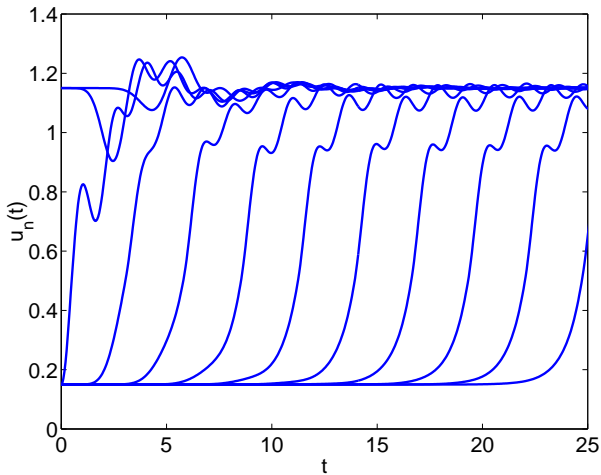}% exi8.eps
\caption{Trajectories $u_n(t)$, $n=8,4,0,-4,...$ when $D=4$,
$\gamma=0.4$, $F=0.15$.}
\label{figura2}
\end{center}
\end{figure}

We will focus on initial kink-like initial states that generate
ordered dynamics: $u_k(t)-{1\over 2}$ changes sign in an ordered way as the 
kink advances. Once the kink has passed, the configuration of the chain 
is close to a shifted
static kink. That is why we use static kinks to obtain simplified
expressions for $u_n(t)$. For instance, let us choose 
a piecewise constant initial profile $u_n^0=F$ for $n\leq 0$ and $u_n^1=1+F$ 
for $n\geq 1,$ with $u_n^1=0$. For $F>0$, Figure \ref{figura2} shows 
that  $u_{-k}(t)-{1\over 2}$ change sign at time $t_k$, $k=0,1,....$, with 
$t_0<t_1<...<t_k<...$. Eventually, the kink may get pinned at some static
configuration and this process stops at some $k$. 
We then use a slightly modified version of the integral expression 
(\ref{st2}) for the static wave fronts to successively remove the integral 
terms in (\ref{st7}) and obtain simple formulae for $u_n(t)$ similar
to (\ref{st3}). In this way, we find a relatively cheap algorithm
for the computation of $u_n(t)$.

Let us describe the algorithm for $F\geq 0$ and an initial step-like 
state $u_n^0$ with $u_0^0<{1\over 2}<u_1^0$, as in Figure \ref{figura2}.
We must distinguish two cases:  $0<F\leq F_{cs}(D)$ and $F>F_{cs}(D).$

\subsubsection{Case $0<F\leq F_{cs}(D)$}   

In this case, the stationary wave fronts can be used to generate
a faster algorithm for obtaining $u_n(t)$.
we remove the integrals in (\ref{st7}) by using the static wave 
front solution of (\ref{e1}), $s_n$, such that $s_0< {1\over 2} < s_{1}$.

{\it Initial stage.}
Formula (\ref{st3}) allows to compute $u_n(t)$ up to the time 
$t_0$ at which $u_0(t)-{1\over 2}$ changes sign.
For $t>t_0$ we compute $u_n(t)$ using as initial
data $u_n(t_0)$ and $u_n'(t_0)$ at $t_0$. The later is obtained
differentiating (\ref{st3}):
\begin{eqnarray}\begin{array}{l}
u_n'(t_0)=\sum_k \Big[{d G_{nk}^0(t_0)\over dt} u_k^1 + 
{d G_{nk}^1(t_0) \over dt} (u_k^0-s_k)\Big].  
\end{array}\label{st8}
\end{eqnarray}
For $t_0<  t \leq t_1$, equation (\ref{st7}) becomes:
\begin{eqnarray} \begin{array}{l}
u_n(t) \!=\!\sum_k \!\Big[G_{nk}^0(t-\! t_0) u_k'(t_0) \!+\! 
G_{nk}^1(t-\! t_0) u_k(t_0)\Big]   \\ 
+  \int_{t_0}^{t}\sum_{k>-1} G_{nk}^0(t -z)  dz 
\!+\! F  \int_{t_0}^{t}\sum_{k} G_{nk}^0(t -z)  dz
\end{array}\label{st9}
\end{eqnarray}
Now, $u_{-1}(t_0) < {1\over 2} < u_0(t_0)$ and we must use the shifted
stationary solution $v_n=s_{n+1}$, that satisfies $v_{-1}< {1\over 2}
< v_{0}$. Observing that $s_{n+1}$ solves (\ref{e1}) with initial data
$s_{n+1},0$ at time $t_{0}$ we obtain the formula:
\begin{eqnarray}\begin{array}{l}
s_{n+1} \!=\!\sum_k  G_{nk}^1(t-t_{0}) s_{k+1}
+  F  \int_{t_{0}}^{t}\sum_{k} G_{nk}^0(t -z)  dz \\
+  \int_{t_{0}}^{t}\sum_{k>-1} G_{nk}^0(t -z)  dz  
\end{array}\label{st10}
\end{eqnarray}
Subtracting  (\ref{st10}) from (\ref{st9}) we find:
\begin{eqnarray}\begin{array}{l}
u_n(t)= s_{n+1} + \sum_k G_{nk}^0(t-t_0) u_k'(t_{0})  \\ 
+ \sum_k G_{nk}^1(t-t_0) (u_k(t_{0})-s_{k+1}),  
\end{array}\label{st11}
\end{eqnarray}
up to the time $t_1$ at which $u_{-1}(t)-{1\over 2}$ changes sign.

{\it Generic step.}
Once we have computed the time $t_l$ at which
$u_l(t)-{1\over 2}$ changes sign, we calculate the new initial
data $u_n(t_l)$ and $u_n'(t_l)$:
\begin{eqnarray}\begin{array}{r}
u_n(t_l)= s_{n+l}+ \sum_k  G_{nk}^0(t_l-t_{l-1}) u_k'(t_l)  \\
+ \sum_k G_{nk}^1(t_l-t_{l-1})  (u_k(t_l)-s_{k+l}), 
\end{array}\label{st125}\\
\begin{array}{r}
u_n'(t_l)=\sum_k {d G_{nk}^0\over dt}(t_l-t_{l-1}) u_k'(t_l) + \\
\sum_k {d G_{nk}^1\over dt}(t_l-t_{l-1})  (u_k(t_l)-s_{k+l}) .
\end{array}\label{st13}
\end{eqnarray}
Then the evolution of the chain for $t> t_l$ is given by the formula:
\begin{eqnarray}\begin{array}{r}
u_n(t)= s_{n+l+1} + \sum_k G_{nk}^0(t-t_l) u_k'(t_l) + \\ 
\sum_k G_{nk}^1(t-t_l) (u_k(t_l)-s_{k+l+1}),  
\end{array} \label{st12} \end{eqnarray}
until either $u_{-(l+1)}(t)-{1\over 2}$ or  $u_{-l}(t)-{1\over 2}$ change sign.
If  $u_{-(l+1)}(t)-{1\over 2}$ changes its sign at a time $t_{l+1}$,
we start a new step using $s_{n+l+1}$ to compute $u_n(t)$.
If  $u_{-l}(t)-{1\over 2}$ reverses its sign at a time $t_{l+1}$,
we start a new step using $s_{n+l-1}$ to compute $u_n(t)$.

\subsubsection{Case $F> F_{cs}(D)$}   

In this case, it is convenient to remove the integral in Eq.\ (\ref{st7}) by
using as $s_n$ the static wave front solution of (\ref{e1}) corresponding to 
 an applied force $F=F_{cs}(D)$, and such that $s_0< {1\over 2} < s_{1}$.
Recall that  there are no stationary wave fronts for $F>F_{cs}$.

{\it Initial stage.} Subtracting (\ref{st2}) at $F_{cs}(D)$
from (\ref{st1}) we find:
\begin{eqnarray} \begin{array}{l}
u_n(t)=s_n+\sum_k G_{nk}^0(t) u_k^1  
+\sum_k G_{nk}^1(t) (u_k^0-s_k) \\
+  (F-F_{cs})  \int_{0}^{t}\sum_{k} G_{nk}^0(t-z)  dz
\end{array} \label{st14}
\end{eqnarray}
The remaining integral term can be removed by observing that $1$ is 
a solution of (\ref{e1}) with $F=0$ and initial data $u_n(0)=1,$
$u_n'(0)=0$:
\begin{eqnarray} 
1=\sum_k  G_{nk}^1(t)  
+ \int_{0}^{t}\sum_{k} G_{nk}^0(t-z)  dz
\label{st15}
\end{eqnarray}
Multiplying (\ref{st15}) by $(F-F_{cs})$ and inserting the result in 
(\ref{st14}) we obtain:
\begin{eqnarray} \begin{array}{r}
u_n(t)=s_n+(F-F_{cs})+\sum_k G_{nk}^0(t) u_k^1 \\ 
+\sum_k G_{nk}^1(t) (u_k^0-s_k-(F-F_{cs})) 
\end{array} \label{st16}
\end{eqnarray}
up to the time $t_0$ at which $u_0(t)-{1\over 2}$ changes sign. 
For $t> t_0$:
\begin{eqnarray}\begin{array}{l}
u_n(t)= s_{n+1}+(F-F_{cs}) +\sum_k G_{nk}^0(t-t_0) u_k'(t_0) \\ 
+\sum_k G_{nk}^1(t-t_0) (u_k(t_0)-s_{k+1}-(F-F_{cs})), \\
u_n'(t_0)\!=\!\!\sum_k\!\! \Big[{d G_{nk}^0(t_0)\over dt} u_k^1 \!+\! 
{d G_{nk}^1(t_0) \over dt} (u_k^0\!-\!s_k\!-\!F\!+\! F_{cs}\!)\!\Big],  
\end{array}\label{st17}
\end{eqnarray}
up to the time $t_1$ at which $u_{-1}(t)-{1\over 2}$ changes sign.

{\it Generic step}. Similar to the generic step for $F\leq F_{cs}$
but replacing $s_n$ by $s_n+(F-F_{cs})$.

\subsubsection{Numerical implementation}

We will use (\ref{st12})-(\ref{st13}) and (\ref{st17}) to study
the dynamics of the chain in Section \ref{sec:traveling}.
Due to translational invariance $G_{nk}^0=G_{n-k,0}^1$ and 
$G_{nk}^1=G_{n-k,0}^1$. To calculate $u_n(t)$, 
we only need to compute $G_{n0}^0(t)$, $G_{n0}^1(t)$ for a time interval 
 $[0,T]$, $T\leq {\rm max}_l |t_{l+1}-t_l|$ and for $|n|\leq N$, where $N$
is sufficiently large. We calculate the integrals $G_{n0}^0(t)$, 
$G_{n0}^1(t)$, ${d G_{n0}^0(t)\over dt}$ and ${d G_{n0}^1(t)\over dt}$ by 
means of the Simpson rule, choosing a step smaller than the period
of the oscillatory factors.
The value $N$ is selected so as to make the error introduced by the
truncated series $\sum_{|n-k|\leq N}$ sufficiently small. 
This is possible because the Green functions and their derivatives decay
as $|n-k|$ grows.

A more general version of our algorithm will be presented elsewhere 
\cite{preprint}.

\section{Stability of traveling kinks}
\label{sec:traveling}

In this section we introduce a strategy to study the stability 
of traveling wave fronts in (\ref{e1}). 

Traveling wave fronts are constructed  by inserting
$w_n(t)=w(n-ct)$ in (\ref{e1}) to produce a nonlinear eigenvalue 
problem for the profile $w(x)$ and the speed $c$. Assuming $w(x)<{1\over 2}$
for $x<0$ and $w(x)>{1\over 2}$ for $x>0$, the problem becomes linear.
The wave profiles are computed as contour integrals, imposing
$w(0)={1\over 2}$ to find a relationship between $c$ and $F$ \cite{atk65,wavy}. 
The law $F(c)$ and the shape of the wave profiles are controlled by 
the poles contributing to the contour integrals. The relevant poles
depend on the strength of the damping. For large damping, we have
complex poles with large imaginary parts. The dependence law
$F(c)$ is monotonically increasing and the wave profiles are 
monotone. For small damping, poles with small imaginary parts
become relevant, in increasing number as the speed $c$ decreases.
The function $F(c)$ oscillates for small speeds. Different
oscillatory wave profiles with different speeds may coexist for 
the same $F$. At zero damping, those poles become real and the
wave profiles develop non decaying oscillations. For some ranges
of speeds, the waves constructed in this way violate the restriction
$w(x)<{1\over 2}$ for $x<0$ and $w(x)>{1\over 2}$ for $x>0$. Those ranges should
be investigated with a modified technique allowing for a finite
number of turning points. 

Complex variable methods yield families of explicit 
wave solutions but give no information on their stability. 
Numerical tests \cite{wavy} and physical context \cite{atk65} 
suggest the stability of traveling kinks that have monotone
leading edges and large enough speeds.
Figures (\ref{figura3})-(\ref{figura5}) depict the wave profiles 
for decreasing $\gamma$. We now confirm that these wave fronts
are stable.
The travelling wave $w_n(t)$ is stable for the dynamics of the
chain when the solutions $u_n(t)$ of (\ref{e1}) remain near 
$w_n(t)$ for all $t>0$ if the initial states $u_n^0$, $u_n^1$
are chosen near $w_n(0),w_n'(0)$. 
Controlling the evolution of $u_n(t)$ is more or less difficult
depending on the properties of the Green functions.
We distinguish two cases: positive Green functions (large damping)
and oscillatory Green functions (small damping).

\subsection{Strong damping}
\label{sec:strong}

For large damping $\gamma^2 \gg 4 $, we know that the wave front 
profiles are monotonically increasing and that the Green functions 
are positive and decay exponentially in time (cf.\ Appendix B). The main 
result of this section is the following stability theorem, whose
proof can be found in Appendix C:

{\sc Theorem.}
{\it Let us select the wave front profile so that $w_n(t)=w(n-ct -{1\over 2})$, with
$c<0$, and $F>0$. If we choose the initial states for (\ref{e1}), $u_n^1$ 
and $u_n^0$, in the set:
\begin{eqnarray}
w_n(-\tau)< u_n^0<w_n(\tau), \quad
0< \tau \ll {1\over 2|c|} \label{tw01}  \\
\begin{array}{l}
|w_n'(-\tau)-u_n^1| \ll u_n^0-w_n(-\tau)\\
|w_n'(\tau)-u_n^1| \ll w_n(\tau)-u_n^0
\end{array} \label{tw02}
\end{eqnarray}
then
\begin{eqnarray}
w_n(t-\tau)< u_n(t) < w_n(t+\tau)   \label{tw03}
\end{eqnarray}
for all $n$ and $t>0$.}

In other words, if the initial oscillator configuration is sandwiched 
between two wave front profiles with different phase shifts, $w_n(-\tau)$ and 
$w_n(\tau)$, (with a sufficiently small $\tau$), then the oscillator chain 
remains trapped between the two shifted profiles $w_n(t-\tau)$ and
$w_n(t-\tau)$ forever, provided $|u_n^1-w_n'(0)|$ is sufficiently small. This 
implies the dynamical stability of the wave. The more involved argument
explained in Section \ref{sec:conser} for conservative dynamics can be
used to prove that the wave fronts are also asymptotically stable. 

\begin{figure}
\begin{center}
\includegraphics[width=8cm]{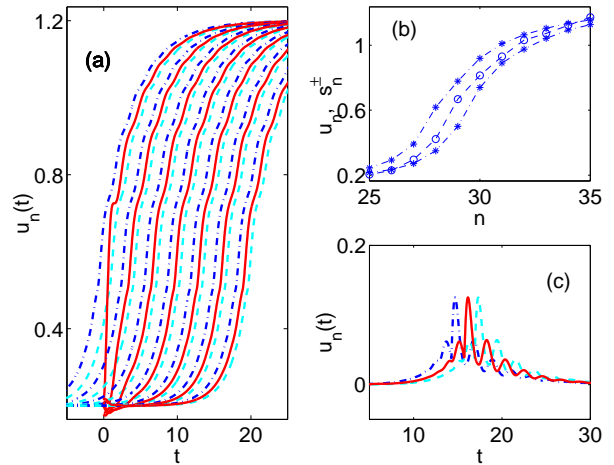}% exi2.eps
\caption{For  $ D=4$, $\gamma=2.2$, and $F=0.2$: 
(a) Compared time evolution of $w_n(t+\tau)$ (dot-dashed line)
$w_n(t-\tau)$ (dashed line) and $u_n(t)$  (solid line)
when $n=0,-1,-2,...$,
(b) Compared profiles $u_n(T)$ (circles), $w_n(T\pm \tau)$ (asterisks),
(c) Compared time evolution of $w_n'(t\pm \tau)$ (dot-dashed
and dashed lines) and $u_n'(t)$ (solid line).
}
\label{figura3}
\end{center}
\end{figure}

\begin{figure}
\begin{center}
\includegraphics[width=8cm]{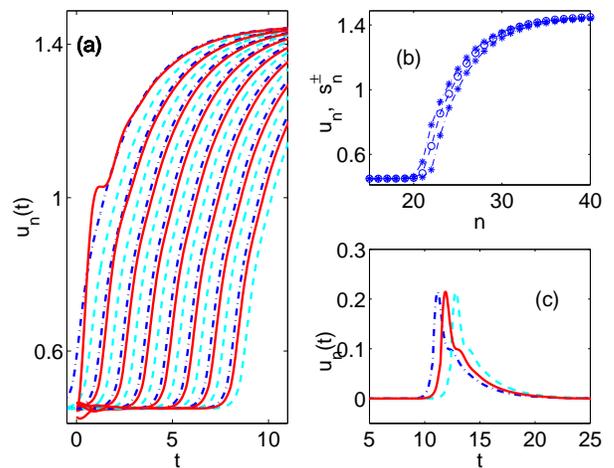}% exi3.eps
\caption{Same as Figure \ref{figura3} when $F=0.45$. }
\label{figura4}
\end{center}
\end{figure}

Furthermore, the basin of attraction of a particular traveling wave
is larger than (\ref{tw01}) - (\ref{tw02}), as shown in Figures 
\ref{figura3}-\ref{figura4} for $F >F_{cs}(D)$. The initial oscillator
configuration in this figure is a step function, $u_n^0=F$ for $n\leq 0$ and
$u_n^0=1+F$ for $n\geq 1$,
with a superimposed small random disturbance. The initial velocity profile 
fluctuates randomly about zero with a small amplitude.
After an initial transient, the trajectories get trapped between advanced 
wave fronts, $w_n(t+\tau)$, and delayed wave fronts, $w_n(t-\tau)$. 
Moreover, they converge to a shifted wave front, $w_n(t+\alpha)$, as 
$t\rightarrow \infty$.

\subsection{Conservative dynamics}
\label{sec:conser}

\begin{figure}
\begin{center}
\includegraphics[width=8cm]{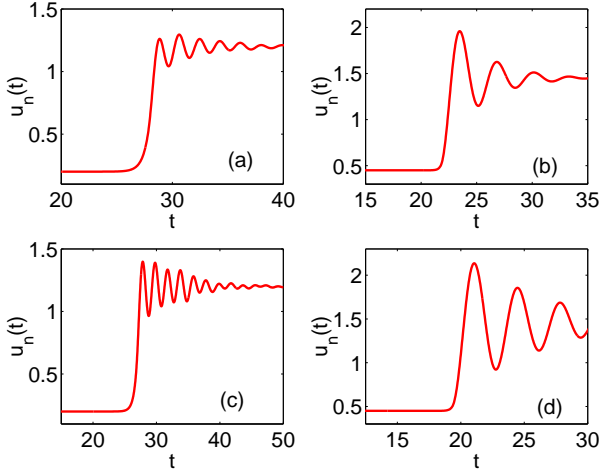}% exi5.eps
\caption{Trajectories $u_{n}(t)$
when $D=4$: (a) $\gamma=0.2$, $F=0.2$, 
(b) $\gamma=0.2$, $F=0.45$, (c) $\gamma=0.1$, $F=0.2$,
(d) $\gamma=0.1$, $F=0.45$.}
\label{figura5}
\end{center}
\end{figure}

For small damping  $\gamma^2 \ll 4$, we know that the kink profiles 
develop oscillations in the trailing edge (see Figure \ref{figura5})
and that the Green functions oscillate and change sign (cf.\ Appendix B).  
However, $G_{nk}^0(t)$ and $G_{nk}^1(t)$ are positive for $0\leq t 
\leq T^*=T(\gamma,D)$. This critical time $T^*$ plays a key role for 
the stable propagation of waves.
We will show in this section that kinks are stable provided
$|c|>{1\over T^*}.$ Our argument does not say anything about the
stability of kinks with lesser speeds.
Moreover, $T^* \rightarrow 0$ and  our lower bound on the wave front
velocity tends to infinity, in the continuum limit. 

\begin{figure}
\begin{center}
\includegraphics[width=8cm]{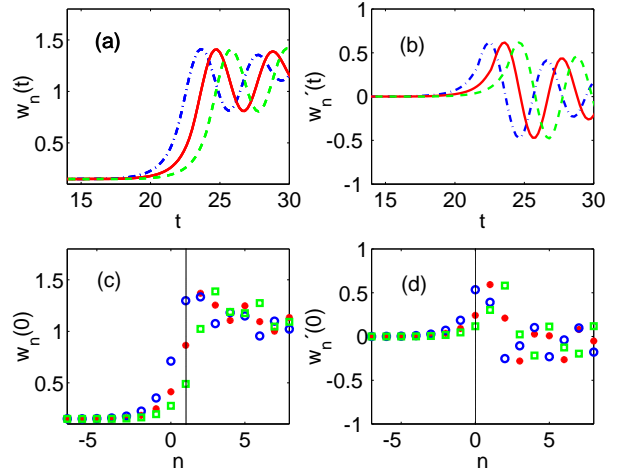}%  
\caption{ (a) Trajectories $w_{n-1}(t)$ (dash-dotted), $w_{n}(t)$
(solid), $w_{n+1}(t)$ (dashed); (b) same for $w_{n-1}'(t)$, $w_{n}'(t)$,
$w_{n+1}(t)$; (c) initial configurations for $w_n(\tau)$ (circles), 
$w_n(0)$ (asterisks), $w_n(-\tau)$ (squares), the vertical line defines 
$n_0$; (d) same for $w_n'(\tau)$, $w_n'(0)$, $w_n'(-\tau)$, the vertical
lines defines $n_1$.}
\label{figura6}
\end{center}
\end{figure}

We show in Appendix B that a rough estimate for $T^*$ is provided 
by  ${2 \pi   \over \sqrt{4(1+4D)-\gamma^2}}$.
For $\gamma=0$ and  $D=4$, as chosen in our Figures 
(\ref{figura5})-(\ref{figura6}),
$T^*> 1$. Then, kinks with  $|c|> 1$ are stable. 
In \cite{atk65} and \cite{wavy}, stability was conjectured for 
speeds larger than the last minimum of  $F(c)$, which is 
attained at $c_{cd}\sim 0.74$. 

For small or zero damping we cannot use the previous comparison
arguments because the trailing edge of the  traveling wave front
oscillates and monotonicity does not hold there. If we  look
at the traveling wave front profiles, it becomes clear that we 
should compare the monotone leading edges of the fronts.
Figure \ref{figura6} (a)-(b) depicts the trajectories $w_n(t)$ and their 
time derivatives $w_n'(t)$ for a particular traveling wave front. We 
observe that $w_{n-1}(t)< w_n(t) < w_{n+1}(t)$ and $w_{n-1}'(t) 
< w_n'(t) <w_{n+1}'(t)$ up to a certain time.  
Figure \ref{figura6} (c)  shows the initial configurations for $w_n(0)$ 
and the shifted waves $w_n(-\tau)$, $w_n(\tau)$.
$w_n(0)$ is sandwiched between $w_n(-\tau)$ and $w_n(\tau)$ up to a
point $n_0$. Figure \ref{figura6} (d) depicts the initial velocity
profiles $w_n'(0)$, $w_n'(-\tau)$ and $w_n'(\tau)$.
$w_n'(0)$ is sandwiched between $w_n'(-\tau)$ and $w_n'(\tau)$ up to a
point $n_1$. $n_0$ and $n_1$ mark the onset of the oscillatory tails.
In general, $0 \leq n_0\leq n_1$. As the wave advances, the ranges
of $n$ for which  $w_n(t-\tau)<w_n(t)=w(n-ct-{1\over 2})<w_n(t+\tau)$ change 
with $t$.

The main result of this section is the following stability theorem, 
whose proof can be found in Appendix D:

{\sc Theorem.}
{\it Let us select the wave front profile so that $w_n(t)=w(n-ct-{1\over 2})$, 
with $c<0$ and $F>0$. Let $T^*$ be the maximum time up to which
the Green functions $G_{nk}^0(t)$ and $G_{nk}^1(t)$ remain positive.
We assume that the speed  $|c|>{1\over T^*}$ and
choose the initial states for (\ref{e1}), $u_n^1$ 
and $u_n^0$, in the set:
\begin{eqnarray}
\begin{array}{ll}
w_n(-\tau)< u_n^0<w_n(\tau),\; n\leq n_0
\quad 0< \tau \ll {1\over 2|c|}\\
w_n'(-\tau)< u_n^1<w_n'(\tau),\; n\leq n_1
\end{array}\label{tw04} \\ 
\begin{array}{l}
\sum_{n} |u_n^0-w_n(0)| < \epsilon, \quad 
\sum_{n} |u_n^1-w_n'(0)| < \epsilon.
\end{array}\label{tw05}
\end{eqnarray}
for $\epsilon>0$ small and $\tau< \epsilon$.
Then, we can find an increasing sequence of times $t_k$, $k=0,1,...$,
with $t_{-1}=0$, such that:
\begin{eqnarray}
\begin{array}{l}
w_n(t-\tau)< u_n(t) < w_n(t+\tau),\quad  n\leq n_0-k\\
w_n'(t-\tau)< u_n'(t) < w_n'(t+\tau),\quad n\leq n_1-k
\end{array}\label{tw06}
\end{eqnarray}
for  $t_{k-1}\leq t<t_k$. Furthermore, for $t>0$
and any $n$, we have:
\begin{eqnarray}
|u_n(t)-w_n(t)| \leq \sum_k |G_{nk}^1(t)| |u_k^0-w_k(0)| \nonumber \\
+ \sum_k |G_{n0}^0(t)||u_k^1-w_k'(0)|+C(t) \label{tw07} \\
C(t)= \sum_{k\leq 0, t\geq T_k-\tau}
\int_{T_k-\tau}^{{\rm Min}(T_k+\tau,t)}   G_{nk}^0(t-z) dz,  
\nonumber 
\end{eqnarray} 
in which $T_k={k\over |c|} +{1\over 2|c|}$. Thus, the traveling
wave front is stable when $\gamma=0$ or asymptotically stable
when $\gamma>0$.}

Let us clarify the meaning of (\ref{tw07}). For $\gamma>0$,
the sums  $\sum_k |G_{nk}^1(t)| $,  $\sum_k |G_{nk}^0(t)|$
decay exponentially with time. For small $\tau$, the function 
$|C(t)| \sim 2 \tau \sum_{k\leq 0, t\geq T_k-\tau} |G_{nk}^0(t-T_k)|$.
This sum is finite and decays with time. This explains our asymptotical 
stability claim. When $\gamma=0$, the sums  $\sum_k |G_{nk}^1(t)| 
|u_k^0-w_k(0)| $,   $\sum_k |G_{nk}^0(t)| |u_k^0-w_k(0)|$  are bounded
by a constant times  ${\rm Max}_k  |u_k^0-w_k(0)|+ {\rm Max}_k 
|u_k^0-w_k(0)|.$ The function $|C(t)|$ is bounded by a constant
times $\tau$ and is made small be choosing $\tau$ small. This explains
our stability claim in the conservative case.

The inequalities (\ref{tw06}) tell us that the leading edge
of the propagating kink is sandwiched between the leading edges
of the shifted traveling wave fronts $w_n(t+\tau)$ and $w_n(t-\tau)$. 
As the kink $u_n(t)$ advances, the times $t_k$
at which $u_{-k}(t)-{1\over 2}$ changes sign are bounded by the
times at which the advanced and delayed wave fronts cross
${1\over 2}$: $T_k-\tau \leq t_k \leq T_k+ \tau$. This fact is the key
for obtaining the stability bound (\ref{tw07}).

\section{Coexistence}
\label{sec:coex}

The results in Section \ref{sec:conser} indicate that stable
static and traveling kinks may coexist. The only restriction 
on the traveling kinks is  the monotonicity of the leading
edge and a low bound on the speed. These conditions are satisfied 
by traveling wave fronts for a range of forces in which static 
wave fronts also exist.
We show in this section how oscillating Green functions may force 
initial kink-like configurations (which would be pinned for large 
damping) to evolve into a traveling wave fronts provided the
damping is small enough.

We fix $F<F_{cs}$ and select the static kink  $s_n$  
constructed in Section \ref{sec:static} for (\ref{e1})  with
$s_0<{1\over 2}<s_1$. Let the initial condition for 
(\ref{e1}) be a piecewise constant   profile: 
$u_n^0=F$ for $n\leq 0$, $u_n^0=1+F$ for $n\geq 1$ and $u_n^1=0$.
Let $T^*$ the maximum time up to which $G_{nk}^0$ and $G_{nk}^1$
remain positive.

As long as $u_n(t)-{1\over 2}$ does not change sign for any $n$,
$u_n(t)$ is given by  formula (\ref{st2}) in Section \ref{sec:static}.
We have $\sum_k G_{nk}^0(t) u_k^1\geq 0$ for $t\leq T^*$.
Initially, $G_{nk}^1(t)$ is concentrated at $k=n$ and the 
sign of $\sum_k G_{nk}^1(t) (u_k^0-s_k) $ is decided
by the sign of $u_n^0-s_n$.
If $u_n^0>s_n$, $u_n(t)\geq s_n$. In our
case, this is true for $n\geq 1$.
If $u_n^0<s_n$, $u_n(t)$ increases towards $s_n$
as  $\sum_k G_{nk}^1(t) (u_k^0-s_k) $ decays.
By our choice of the initial state, $u_0(t)$ grows faster
than the other components $u_n(t)$, $n<0$.

Now there are two possibilities depending on the value of the
damping coefficients. For large damping, $G_{nk}^1(t)$ is
positive for all times and $G_{nk}^0$ decays fast. 
Then $u_0(t)$ cannot surpass $s_0$.
This initial data is pinned for large damping. 

For small values of the damping,  $G_{nk}^1(t)$ changes sign. 
Then $u_0(t)$ given by (\ref{st2}) may surpass $s_0$ and ${1\over 2}$ 
since the term $\sum_k G_{nk}^1(t) (u_k^0-s_k) $ becomes
positive for $t\geq T^*$.  This process can be iterated
to get a stably propagating wave, see Figure \ref{figura0}. 
A prediction for the speed is found in this way:
it is the reciprocal of the time that 
$s_0+G_{00}^1(t)(F-s_0)$ needs to reach ${1\over 2}$.

\section{More general potentials}
\label{sec:potential}

We have focused our study on periodic piecewise parabolic potentials
$V(u)={u^2 \over 2}$, $|u|<{1\over 2}$. For these potentials,  
families of static and traveling wave fronts can be constructed
analytically.  Schmidt \cite{sch79} and later authors \cite{bre97,fla99} 
found exact monotone wave fronts of conservative systems by constructing 
models with nonlinearities such that the desired wave fronts were solutions 
of the models. For damped Frenkel-Kontorova or quartic double well potentials,
stably propagating wave fronts have been found numerically \cite{wavy}. 
Numerical studies of kink propagation in the conservative Frenkel-Kontorova 
model were carried out in \cite{pey84}. 

The stability of propagating kinks in these models can be studied
adapting the methods developed in this paper, but the analysis is
technically more complicated \cite{preprint}.
For instance, taking $V(u)={u^2 \over 2} $ for $|u|<{1\over 2}$, 
${1\over 4}-{(u-1)^2\over 2}$ for $|u-1|<{1\over 2}$ we get a {\em continuous}
piecewise linear source 
\begin{eqnarray}
g(s)= \left\{\begin{array}{ll}
s  \quad s < 1/2\\
-s+{1\over 2} \quad {1\over 2} < s < 3/2 \\
s - 1 \quad s \geq 3/2 \\
\end{array}\right. \nonumber
\end{eqnarray}
The arguments in Sections \ref{sec:strong} and  \ref{sec:conser} 
can be adapted by including   new terms  $ \int_{0}^{t}
\!\!\sum_{ {1\over 2}< u_k< {3\over 2}} G_{nk}^0(t-z) (2u_k(z)-{1\over 2}) dz $  
in the integral expressions (\ref{tw1})-(\ref{tw2}) and  using
that the function $h(s)=2s-1/2$ is increasing in
$1/2< s< 3/2$. Similarly, for a Frenkel-Kontorova potential,
we write $g(s)=-as + (\sin(s)+as), a>0$. Then, we find the integral 
expression (\ref{ap8}) with a nonlinear source $f_k=-\sin(u_k)+au_k+F$,
using the Green functions for the linear operator
$ u_n''+ \gamma u_n' - D(u_{n+1}-2u_n+u_{n-1})+ au_n$.
 The parameter $a$ is chosen to ensure
adequate monotonicity properties for $h(s)=-\sin(s)+as$ \cite{preprint}.

\section{Conclusions}
\label{sec:conclusions}

We have developed a nonlinear stability theory for wave fronts in 
conservative and damped Frenkel-Kontorova models with 
piecewise linear sources based on integral formulations.
 Our results provide an analytical 
basis for the distinction between static and dynamic Peierls 
stresses, which arise as thresholds for the existence
of stable static and traveling wave fronts.
With little or zero damping, stable propagation of fronts is possible
when their speeds surpass a critical value. The corresponding
wave front profiles have a monotone leading edge, and, possibly,
an oscillatory wake. Wave fronts can be oscillatory
and yet stable. Whether slow  wave fronts  showing oscillations 
in the leading and trailing edges are stable remains an open 
question \cite{wavy}. It is remarkable that high  
order quasicontinuum approximations such as those by 
Rosenau \cite{rosenau} or by Boussinesq \cite{boussi} have wave 
solutions comparable to the fast waves of the discrete conservative  
model \cite{truski}.

Together with the stability theory we have presented an 
algorithm for the numerical computation of the dynamics
of kinks. Our scheme has good stability properties and
avoids distortions originated by artificial boundary 
conditions and time discretization.

\acknowledgments
This work has been supported by the Spanish MCyT through grant
BFM2002-04127-C02, and by the European Union under grant 
HPRN-CT-2002-00282. The author thanks Prof. L.L. Bonilla for
a careful reading of the paper.

\appendix
\section{Green functions}
\label{sec:green}

We want to find an integral representation of the solution of the
problem:
\begin{eqnarray}
  u_n''+ \gamma u_n' = D(u_{n+1}-2u_n
+u_{n-1})- u_n +f_n  \label{ap1} \\
u_n(0)=u_n^0, \quad u_n'(0)=u_n^1 \label{ap2}
\end{eqnarray}
with $\gamma \geq 0, D>0$. Firstly, we get rid of the 
difference operator by using the generating functions
$p(\theta,t)$, $f(\theta,t)$:
\begin{eqnarray}
p(\theta,t)= \sum_n u_n(t) e^{-i n\theta}, \;
f(\theta,t)= \sum_n f_n(t) e^{-i n\theta} \label{ap3}
\end{eqnarray}
Differentiating $p$ with respect to $t$ and using (\ref{ap1}), we 
see that $p$ solves the ordinary differential equation:
\begin{eqnarray}
 p''(\theta,t)+ \gamma p'(\theta,t)+ \omega(\theta)^2 p(\theta,t) 
= f(\theta,t) \label{ap4}
\end{eqnarray}
where $\omega(\theta)^2=1+4D\sin^2({\theta\over 2})$  and the
obvious initial conditions for $p(\theta,t)$ follow from those for
$u_n(t)$.

The  solution $p$ depends on the roots of the polynomial
$ r^2 +\gamma r + \omega(\theta)^2 =0$.
When ${\gamma^2 \over 4}>\omega^2(\theta)$,
\begin{eqnarray}
p(\theta,t)= p(\theta,0) {e^{r_-(\theta)t} r_+(\theta) 
- e^{r_+(\theta)t} r_-(\theta) \over r_+(\theta)-r_-(\theta) } 
\nonumber \\
+ p'(\theta,0) {e^{r_+(\theta)t}-e^{r_-(\theta)t} 
\over r_+(\theta)-r_-(\theta)}
\nonumber \\
+ \int_0^t {e^{r_+(\theta)(t-s)}-e^{r_-(\theta)(t-s)} 
\over r_+(\theta)-r_-(\theta)} 
f(\theta,s)  ds
\label{p1}
\end{eqnarray}
with 
$r_{\pm}(\theta)={-\gamma \pm \sqrt{\alpha(\theta)} \over 2}<0$
and $\alpha(\theta)= \gamma^2 -4\omega(\theta)^2$.
When ${\gamma^2 \over 4}<\omega^2(\theta)$, the roots are complex:
\begin{eqnarray}
p(\theta,t)= p(\theta,0) e^{{-\gamma \over 2} t} 
\Big[ \cos\big(I(\theta) t\big) +
{\gamma \sin\big(I(\theta)t\big) \over 2 I(\theta) } 
\Big] \nonumber \\
+ p'(\theta,0) \; e^{{-\gamma \over 2} t} 
{\sin\big(I(\theta)  t\big) \over I(\theta)} \nonumber\\
+ \int_0^t e^{{-\gamma \over 2} (t-s)} 
{\sin\big(I(\theta)  (t-s)\big) \over I(\theta)}
f(\theta,s)  ds
\label{p2}
\end{eqnarray}
where $I(\theta)={\sqrt{-\alpha(\theta)} \over 2}$.
When ${\gamma^2 \over 4}=\omega^2(\theta),$
\begin{eqnarray}
p(\theta,t)= p(\theta,0) e^{{-\gamma \over 2} t} (1+{\gamma \over 2} t)
+ p'(\theta,0) \;  e^{{-\gamma \over 2} t}  t \nonumber\\
+ \int_0^t (t-s) e^{{-\gamma \over 2} (t-s)} 
f(\theta,s)   ds \label{p3}
\end{eqnarray}

The solution $u_n(t)$ of (\ref{ap1}) is recovered from the definition
(\ref{ap3}):
\begin{eqnarray}
u_n(t)= \int_{-\pi}^{\pi} {d\theta \over 2 \pi} e^{i n\theta} 
p(\theta,t). \label{ap5}
\end{eqnarray}
Here, $p(\theta,t)$ is defined by (\ref{p1}) when $\theta \in I_1$: 
\begin{eqnarray}
I_1= \{\theta \in [-\pi, \pi] \;| \; {\gamma^2 \over 4}>\omega^2(\theta)\},
\label{ap6}\end{eqnarray}
by (\ref{p2}) when $\theta \in I_2$: 
\begin{eqnarray}
I_2= \{\theta \in [-\pi, \pi] \;| \;{\gamma^2 \over 4}<\omega^2(\theta)\}.
\label{ap7}\end{eqnarray}
and by (\ref{p3}) when $\theta \in I_3$:
\begin{eqnarray}
P= \{\theta \in [-\pi, \pi] \;| \;{\gamma^2 \over 4}=\omega^2(\theta)\}.
\label{ap7.5}\end{eqnarray}
Notice that $I_1=P=\emptyset$ if $\gamma^2<4 $ and $I_2=P=\emptyset$ 
if $\gamma^2>4 (1+4D)$. $P\neq \emptyset$ only if $4 (1+4D)<\gamma^2<4$
and it then consists of two points.
 
Formula (\ref{ap5}) can be rewritten as:
\begin{eqnarray}
u_n(t)=\sum_k \big[G_{nk}^0(t)  u_k'(0) + 
G_{nk}^1(t) u_k(0)\big] \nonumber \\+  \int_0^t \sum_k 
G_{nk}^0(t-s) f_k(s)  ds \label{ap8}
\end{eqnarray}
where 
\begin{eqnarray}\begin{array}{l}
G_{nk}^0(t)= \int_{-\pi}^{\pi} {d\theta \over 2\pi} 
e^{i(n-k)\theta} g^0(\theta,t) \\
G_{nk}^1(t)= \int_{-\pi}^{\pi} {d\theta \over 2\pi} 
e^{i(n-k)\theta} g^1(\theta,t)
\end{array}\label{ap9}\end{eqnarray}
with
\begin{eqnarray}
g^0(\theta,t)=  \left\{ \begin{array}{ll}
{e^{r_+(\theta)t}-e^{r_-(\theta)t} \over r_+(\theta)-r_-(\theta)}, 
\quad \theta \in I_1 \\
 e^{{-\gamma \over 2} t}  t, \quad \theta \in P \\
e^{{-\gamma \over 2} t} {\sin\big(I(\theta) (t)\big) \over I(\theta)}, 
\quad \theta \in I_2 
\end{array}\right. \label{ap10}\end{eqnarray}
\begin{eqnarray}
g^1(\theta,t)=\left\{\begin{array}{ll}
 {e^{r_-(\theta)t} r_+(\theta)- e^{r_+(\theta)t}   r_-(\theta)  
  \over r_+(\theta)-r_-(\theta) }, \quad \theta \in I_1 \\
e^{{-\gamma \over 2} t} (1+{\gamma \over 2} t), 
\quad \theta \in P  \\
e^{{-\gamma \over 2} t} 
\Big[ \!\cos\big(I(\theta) t\big)\! +\!
{\gamma \sin\big(I(\theta)t\big) \over 2 I(\theta) } \!
\Big], \; \theta \in I_2 
\end{array}\right. \label{ap11}\end{eqnarray}

For conservative chains, $\gamma=0$, $G_{nk}^1={d G_{nk}^0\over dt}$ and:
\begin{eqnarray}
G_{nk}^0(t)=\int_{-\pi}^{\pi} {d\theta \over 2\pi} {e^{i(n-k)\theta}
\over \omega(\theta)} \sin(\omega(\theta)t). 
 \label{consg}
\end{eqnarray}
Green functions for hamiltonian chains were studied in 
\cite{bafalui} and earlier in \cite{schrodinger}.
For overdamped chains, they were computed in \cite{fat98}.

\section{Properties of the Green functions}
\label{sec:props}

\begin{figure}
\begin{center}
\includegraphics[width=8cm]{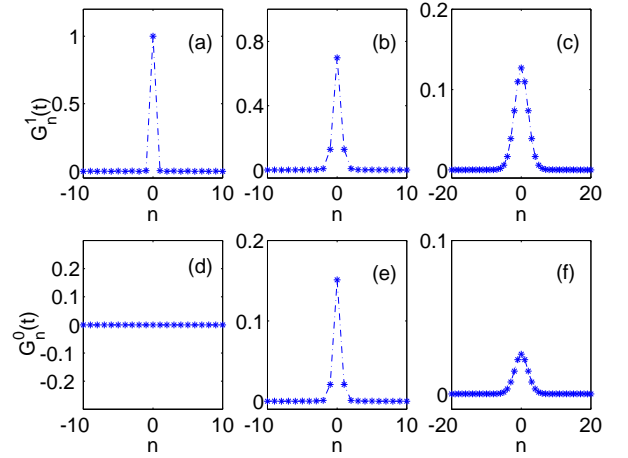}% green1.eps
\end{center}
\caption{Time evolution of the Green functions for 
$D=4$ and $\gamma=10$: 
(a) $t=0$, (b) $t=0.5$, (c) $t=5$; 
(d) $t=0$, (e) $t=0.5$, (f) $t=5$.}
\label{green1}
\end{figure}

\begin{figure}
\begin{center}
\includegraphics[width=8cm]{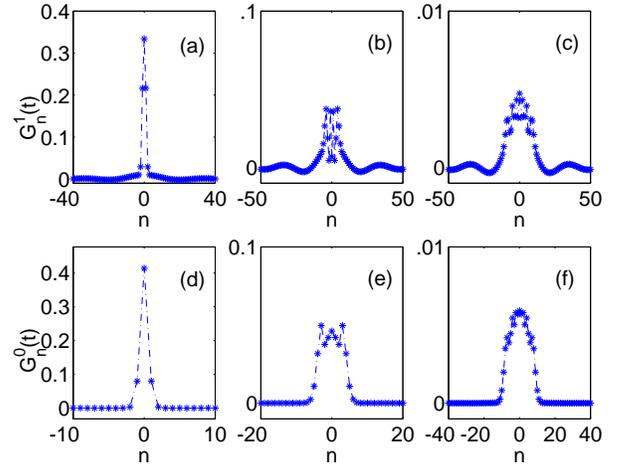}% green4.eps
\end{center}
\caption{Time evolution of the Green functions for 
$D=4$ and $\gamma=2.2$: 
(a) and (d) $t=0.5$, (b) and (e) $t=2.5$, 
(c) and (f) $t=5$.}
\label{green4}
\end{figure}

The Green functions for (\ref{ap1})-(\ref{ap2}) have three
relevant properties: they decay in time, they decay as $|n-k|
\rightarrow \infty$ and are positive for some time.
The property of spatial decay follows from integration by parts
in (\ref{ap9}):
\begin{eqnarray}\begin{array}{l}
G_{nk}^0(t)= {(-1)^l \over  i^l(n-k)^l} 
\int_{-\pi}^{\pi} {d\theta \over 2\pi} 
e^{i(n-k)\theta} {\partial^l g^0(\theta,t) \over 
\partial \theta^l} \\
G_{nk}^1(t)= {(-1)^l \over i^l(n-k)^l}
 \int_{-\pi}^{\pi} {d\theta \over 2\pi} 
e^{i(n-k)\theta} {\partial^l g^1(\theta,t) \over 
\partial \theta^l}
\end{array}\label{prop}\end{eqnarray}
when $n\neq k$. An immediate consequence is that $\sum_k 
|G_{nk}^0(t)|^p$ and $\sum_k |G_{nk}^1(t)|^p$ are finite for 
any $p\geq 1$. Therefore, we may obtain decay results 
as $|n| \rightarrow \infty$ for the solutions $u_n(t)$ of 
(\ref{ap1})-(\ref{ap2}) given by (\ref{ap8}) decay 
when the data $u_n^0$, $u_n^1$, $f_n(t)$ decay.
Figures \ref{green1}-\ref{green4} illustrate the spatial decay 
of $G_n^0(t)= G_{n0}^0(t)$ and $G_n^1(t)=G_{n0}^1(t)$. Notice
that, initially, both are concentrated about $n=0$.

Time decay and positivity depend on the strength of the damping.
Let us start by the {\it strongly damped case: $\gamma^2 > 4(1+4D)$}.
The Green functions are given by (\ref{ap9})-(\ref{ap11})
with $I_2=P=\emptyset$:
\begin{itemize}
\item {\it $G_{nk}^0(t)$ and $G_{nk}^1(t)$ are positive.}
The roots $r_{\pm}(\theta)$ being even with respect to $\theta$,
both $G_{nk}^0(t)$ and $G_{nk}^1(t)$ are real and $e^{i(n-k)\theta}$
can be replaced by $\cos((n-k)\theta)$. The kernels
$g_0(\theta,t)={e^{r_+(\theta)t}-e^{r_-(\theta)t} 
\over r_+(\theta)-r_-(\theta)}$ and 
$g_1(\theta,t)={e^{r_-(\theta)t} r_+(\theta)- e^{r_+(\theta)t}   
r_-(\theta)  \over r_+(\theta)-r_-(\theta) }$ are even, positive, 
reach their maximum values at $\theta=0$ and decay as $\theta$ 
increases to $\pi$. The dominant contribution to the integrals 
(\ref{ap9}) comes from a neighbourhood centered at $\theta=0$, where 
the oscillatory multiplier $\cos((n-k)\theta)$  is positive. 
Thus, both $G_{nk}^0(t)$ and $G_{nk}^1(t)$ are positive. Figure 
\ref{green1} illustrates their evolution as time grows. Notice
the resemblance with the time evolution of heat kernels.
\item $G_{nk}^0(t)$ and $G_{nk}^1(t)$ are bounded uniformly in $n,k$ 
by decaying exponentials in time:
\begin{eqnarray}\begin{array}{l}
|G_{nk}^0(t)|\leq {e^{r_+(0)t}-e^{r_-(\pi)t} \over 
\sqrt{\gamma^2 -4(1+4D)} } \\
|G_{nk}^1(t)|\leq    
{e^{r_-(\pi)t} r_+(\pi)- e^{r_+(0)t}   r_-(0)  
\over \sqrt{\gamma^2 -4 (1+4D)} } 
\label{prop1}\end{array}\end{eqnarray}
\end{itemize}

We come now to {\it intermediate damping $4 <\gamma^2 < 4 (1+4D)$.}
In this case both $I_1$ and $I_2$ are non empty. The piecewise
defined kernels $g_0$ and $g_1$ are still even, take the largest values
near zero (in $I_1$) and the smallest near $\pi$ (in $I_2$).
The dominant contribution to $G_{nk}^0(t)$ and $G_{nk}^1(t)$ comes 
thus from $I_1$ and is positive. This is helped by the fact that
the contribution coming from $I_2$ is initially positive and the
factor $e^{-{\gamma \over 2}t}$ in the oscillatory region $I_2$ 
decays faster than the factor $e^{r_+(\theta)t}$ in the positive 
region $I_1$. Therefore, $G_{nk}^0(t)$ and $G_{nk}^1(t)$ are
essentially positive in this intermediate regime, see Figure
\ref{green4}. This means that their large components are positive,
despite the appearance of some negligible negative components.
They can be roughly bounded by:
\begin{eqnarray}
|G_{nk}^0(t)|\leq C_0 e^{r_+(0)t}, \;
|G_{nk}^1(t)|\leq C_1  e^{r_+(0)t}
\label{prop2} \end{eqnarray}

\begin{figure}
\begin{center}
\includegraphics[width=8cm]{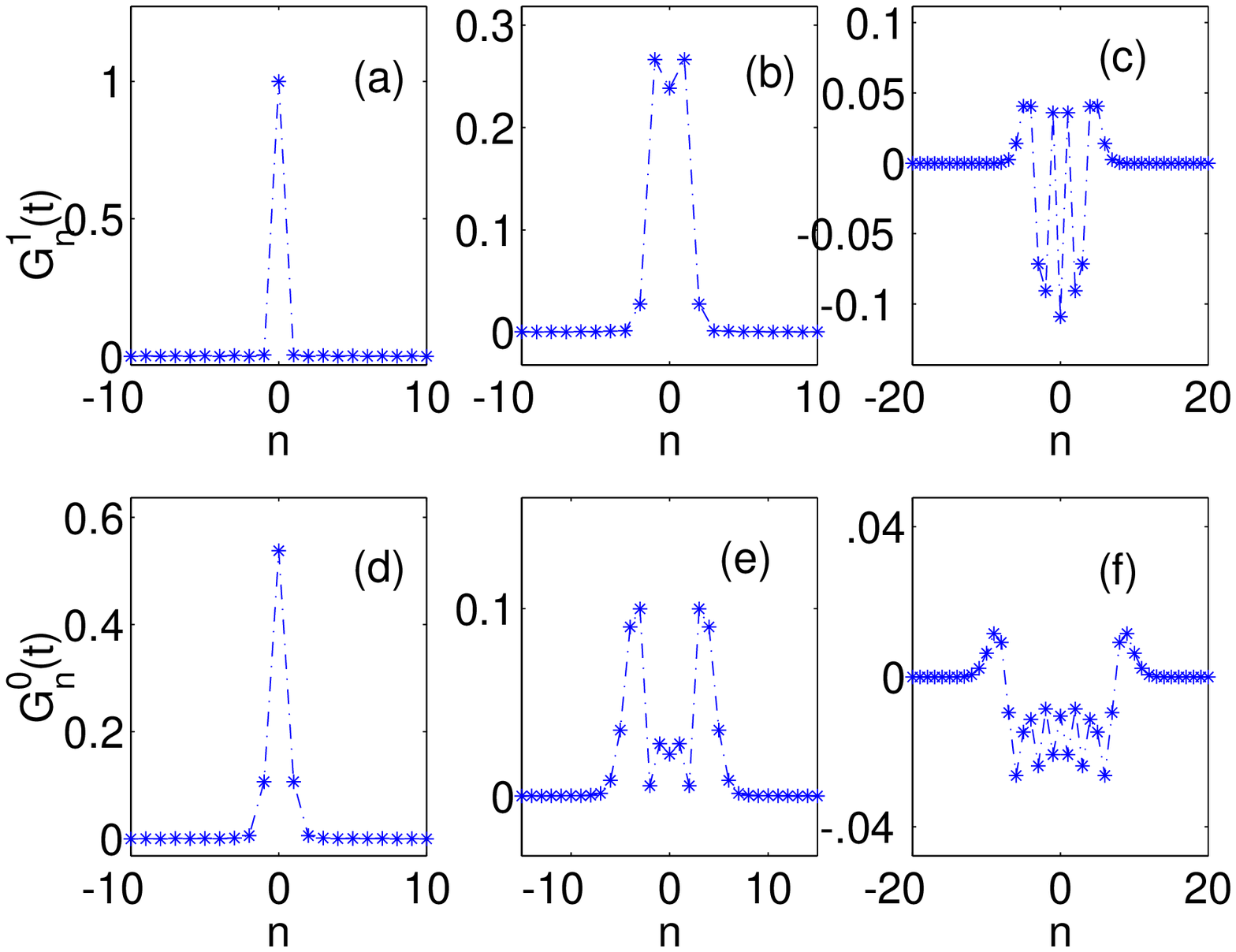}% green2.eps
\end{center}
\caption{Time evolution of the Green functions for 
$D=4$ and $\gamma=1$: 
(a) $t=0$, (b) $t=0.5$, (c) $t=2.5$; 
(d) $t=0.5$, (e) $t=2.5$, (f) $t=5$.}
\label{green2}
\end{figure}
\begin{figure}
\begin{center}
\includegraphics[width=8cm]{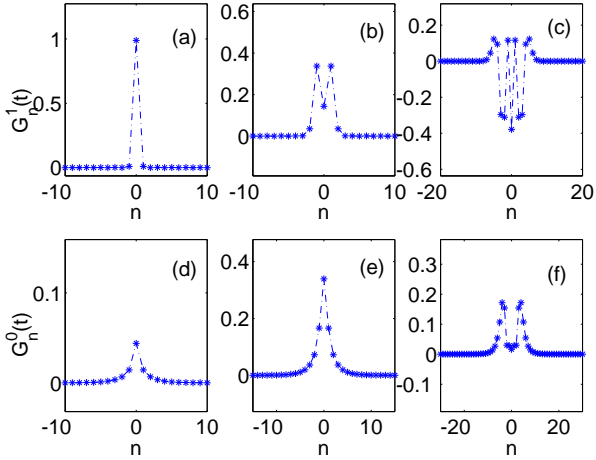}% green3.eps
\end{center}
\caption{Time evolution of the Green functions for 
$D=4$ and $\gamma=0$: 
(a) and (d) $t=0.05$, (b) and (e) $t=0.5$, (c) and (f) $t=2.5$.}
\label{green3}
\end{figure}

We address finally the {\it weakly damped problems with 
$\gamma^2 < 4$.}
In this case, $I_1=P=\emptyset$.
$G_{nk}^0(t)$ and $G_{nk}^1(t)$ are no longer globally positive. 
However, the kernels $g_0(\theta,t)$ and $g_1(\theta,t)$ are  
positive for $|t|<{2\pi \over \sqrt{4(1+4D)-\gamma^2}}=2T$
and $|t|<{\pi \over \sqrt{4(1+4D)-\gamma^2}}=T$, respectively. 
That means that $G_{nk}^0(t)>0$ and $G_{nk}^1(t)>0$ for $t$ in 
those intervals. They remain essentially positive in a larger
interval. The kernels $g_0(\theta,t)$ and $g_1(\theta,t)$ become 
negative for $\theta$ near $\pi$ and remain positive for small $\theta$.
This is enough for the relevant values of $G_{nk}^1(t)$ to remain 
positive up to a critical time $T^*$, often larger than $T$. 
We can get uniform bounds in time:
\begin{eqnarray}\begin{array}{l}
|G_{nk}^0(t)|\leq {2\over \sqrt{4-\gamma^2}} e^{-{\gamma \over 2} t },
\\  |G_{nk}^1(t)|\leq (1+{\gamma \over\sqrt{4-\gamma^2}}
)e^{-{\gamma \over 2} t }
\label{prop3}\end{array}\end{eqnarray}
The same positivity properties and bounds are shared by the Green functions 
in the {\it conservative case $\gamma=0$}. Figures \ref{green2} and 
\ref{green3} illustrate the time evolution of the Green functions.
A  detailed study of the decay properties with respect to $n$ and $t$ 
for conservative problems can be found in \cite{bafalui}.

\section{Stability of traveling wave fronts
for strong damping}
\label{sec:proofstrong}

We now prove  the stability theorem of Section \ref{sec:strong} 
for strong damping. The key idea of the proof is suggested by formula
(\ref{st7}). When we solve (\ref{e1}) starting from different
step-like initial states, we observe three types of terms 
in (\ref{st7}). The second and the third  are increasing functions of 
the step-like configurations. The fourth term does not depend on the 
initial  configuration. The first one can be made small by choosing a 
small velocity profile. 
Our proof proceeds in two steps. First, we establish a few properties
of the traveling wave fronts. Second, we prove the stability bound
(\ref{tw03}).

{\it Step 1: Basic properties of the traveling waves.}
For every $k$, we know that $w_k'(t) >0$. Thus, each $w_k(t)$ crosses
${1\over 2}$ at a definite time $t_k$. Recall that we have selected 
the unique wave profile $w(x)$ satisfying $w(0)={1\over 2}$. Therefore,
$w_{-k}(t)-{1\over 2}$ changes sign at time $T_k={k\over|c|}+{1\over 2|c|}>0$, 
$k=0,1,2,...$. For the shifted waves  $w_k(t+\tau)$ and $w_k(t-\tau)$ the 
changes of sign take place at the shifted times $T_k^{+}=T_k -\tau$ and 
$T_k^{-}=T_k + \tau.$ 

The waves $w_n(t\pm \tau)$  solve the integral equation (\ref{st7}) with 
initial data $w_n(\pm \tau)$ and $w_n'(\pm \tau).$ Using
the  times $T_{k}^{\pm}$, we can rewrite formula (\ref{st7}) 
in a more explicit form:
\begin{eqnarray}
w_n(t\pm \tau)=\sum_k \Big[G_{nk}^0(t) w_k'
(\pm \tau) + G_{nk}^1 w_k(\pm \tau)\Big] 
\nonumber \\ 
+F \!\!\int_{0}^{t}\!\! \sum_{k} G_{nk}^0(t-z)  dz +
\!\!\int_{0}^{t}\! \sum_{k>0} G_{nk}^0(t-z)  dz
\nonumber \\ 
+   \sum_{k\leq 0}\int_{T_k^{\pm}}^{{\rm Max }
(t,T_k^{\pm})} G_{nk}^0(t-z)  dz \label{tw1}
\end{eqnarray}
A term $ \int_{T_k^{\pm}}^{t} G_{n,k}^0(t-z)  dz$ is 
added whenever a factor $w_k(z\pm \tau)-{1\over 2}$ changes sign.

{\it Step 2: Comparing  $u_n(t)$ and $w_{n}(t\pm \tau)$.}
During the initial stage of the evolution of the chain  
$u_0(t)<{1\over 2}<u_1(t)$ and formula (\ref{st7}) reads:
\begin{eqnarray}\begin{array}{r}
u_n(t)=\sum_k \Big(G_{nk}^0(t) u_k^1 + G_{nk}^1(t) u_k^0\Big) +\\ 
\int_{0}^{t}\sum_{k>0} G_{nk}^0(t-z)  dz
+ F \int_{0}^{t}\sum_{k} G_{nk}^0(t-z)  dz
\end{array}\label{tw2}
\end{eqnarray}
 By (\ref{tw01}) and the positivity of $G_{nk}^1(t)$, 
\begin{eqnarray}
\sum_k\!\! G_{nk}^1(t)  w_k(-\tau) \!<\!\!
\sum_k\!\! G_{nk}^1(t)  u_k^0 \!<\!\!
\sum_k\!\! G_{nk}^1(t)  w_k(\tau). 
\label{tw3}
\end{eqnarray}
By (\ref{tw02}),
\begin{eqnarray} 
\begin{array}{l}
\sum_k  G_{nk}^1(t) \big[w_k(\tau)\! -\! u_k^0\big] \! + \! 
G_{nk}^0(t)  \big[w_k'( \tau)\! - \! u_k^1\big] >0\\
\sum_k  G_{nk}^1(t) \big[ u_k^0\! -\! w_k(-\tau)\big]  \! + \!
G_{nk}^0(t)  \big[u_k^1\! - \! w_k'(- \tau)\big] >0
\end{array} \label{tw4}
\end{eqnarray}
Therefore, (\ref{tw1})-(\ref{tw4}) imply
\begin{eqnarray}
w_n(t-\tau)< u_n(t) < w_n(t+\tau),  \label{tw5}
\end{eqnarray}
for all $n$ and $t\leq T_0^+$. Recall that $T_0^+\leq T_0^-$
by definition. Afterwards, $w_0(t+\tau)$ has 
crossed ${1\over 2}$ and $\int_{T_0^+}^{t}G_{n0}^0(t-z)  
dz>0$ must be added in the expression for $w_n(t+\tau)$. The ordering 
(\ref{tw5}) still  holds. At time $T_0^-$, $w_0(t-\tau)$ crosses ${1\over 2}$. 
By (\ref{tw4}), $u_0(t)$ must cross before, at a time $t_0$. 

In this way, we obtain a sequence of times $t_k$ at which $u_{-k}(t)-{1\over 2}$, 
changes sign satisfying $T_k^+ < t_k < T_k^-$, $k=0,1,2,...$. Then,
\begin{eqnarray} 
u_n(t)=\sum_k \Big[G_{nk}^0(t) u_k^1 + G_{nk}^1(t) u_k^0\Big] +
\nonumber \\ 
F \int_{0}^{t}\!\! \sum_{k} G_{nk}^0(t-z)  dz +
\!\!\int_{0}^{t}\! \sum_{k>0} G_{nk}^0(t-z)  dz
\nonumber \\ 
+   \sum_{k\leq 0}\int_{t_k}^{{\rm Max }
(t,t_k)} G_{nk}^0(t-z) dz 
\label{tw6}\end{eqnarray}
and (\ref{tw5}) holds  for all $t$. Our stability claim is proved.

\section{Stability of traveling wave fronts
for conservative dynamics}
\label{sec:proofconser}

In this section, we prove the stability theorem of Section 
\ref{sec:conser} for small or zero damping. 
The notation is the same as in  Appendix \ref{sec:proofstrong} 
and the proof is organized in  two steps.

{\it Step 1: Initial stage.} 
We compare  $u_n(t)$ given by (\ref{tw2}) with the shifted waves 
$w_n(t\pm \tau)$ given by (\ref{tw1}), whereas $u_n'(t)$ is
compared with $w_n'(t\pm \tau)$. The time derivatives are calculated 
by differentiating (\ref{tw1})-(\ref{tw2}). Notice that 
${dG_{nk}^0(t)\over dt} >0$  for small damping when $t\leq T^*$. 
Up to a first critical time  $T_0^+$, $u_k(t)-{1\over 2}$ and
$w_k(t\pm \tau)-{1\over 2}$ keep their sign for all $k$. 
Therefore,
\begin{eqnarray}
\begin{array}{l}
w_n(t\!-\! \tau\!)\! <\! u_n(t) \! <\! w_n(t\! +\! \tau\!), 
\; n\leq n_0, \\
 w_n'(t\!-\! \tau\!)\! <\! u_n'(t) \! <\! w_n'(t\! +\! \tau\!),
\; n\leq n_1, 
\end{array}\label{tw7bis}
\end{eqnarray}
for $t\leq T_0^{+}$.
Recall that, initially,  $G_{nk}^1$ and $G_{nk}^0$, together with 
their derivatives, take on their maximum values for $k$ close to $n$.  
This fact and (\ref{tw03})-(\ref{tw04}) imply:
\begin{eqnarray}
\begin{array}{l}
w_n(t - \tau ) < u_n(t) < w_n(t + \tau), 
\; n\leq n_0-1,\; \\
 w_n'(t-\tau) < u_n'(t)< w_n'(t+\tau),
\; n\leq n_1-1,
\end{array}\label{tw7}
\end{eqnarray}
for  $T_0^+ \leq t \leq T_1^+$, choosing  $\tau<T^*-T_0.$ 
This means that $u_0(t)-{1\over 2}$
changes sign at a time $t_0$ such that $T_0^+<t_0<T_0^- < T_1^+$. 
We then obtain formula (\ref{tw6}) for $u_n(t)$ restricted to
$t\leq T_1^+$. 
By subtracting (\ref{tw1}) from (\ref{tw6}), we find:
\begin{eqnarray}\begin{array}{r}
\sum_n |u_n-w_n |(t) \leq \sum_n |G_{n0}^1(t)| \sum_n 
|u_n-w_n|(0)   \\
+ \sum_n |G_{n0}^0(t)| \sum_n |u_n'-w_n'|(0) +C(t) 
\end{array}\label{tw8}
\end{eqnarray}
where
\begin{eqnarray}
C(t)= \left\{\begin{array}{ll}
0, & t\leq T_0^+, \\
\int_{T_0^+}^{T_0^-}   G_{n0}^0(t-z) dz, &  t > T_0^+,
\end{array} \right.\nonumber
\end{eqnarray}
and
\begin{eqnarray}\begin{array}{r}
\sum_n |u_n'-w_n' |(t) \leq \sum_n |{d G_{n0}^1\over dt}(t)| \sum_n 
|u_n-w_n|(0)   \\
+ \sum_n |{d G_{n0}^0\over dt}(t)| \sum_n |u_n'-w_n'|(0) +R(t), 
\end{array} \label{tw9} 
\end{eqnarray}
where
\begin{eqnarray}
R(t)= \left\{\begin{array}{ll}
0, & t\leq T_0^+, \\
\int_{T_0^+}^{T_0^-} {d G_{n0}^0\over dt}(t-z) dz, &  t > T_0^+,
\end{array} \right.\nonumber
\end{eqnarray}
for $t\leq T_1^+$. Let $S_1={\rm Max}_{[0,\infty)} \sum_n |G_{n0}^1(t)|$,
$S_2={\rm Max}_{[0,\infty)} \sum_n |G_{n0}^0(t)|$,
$S_3={\rm Max}_{[0,\infty)} \sum_n |{dG_{n0}^1\over dt}(t)|$ and
$S_4={\rm Max}_{[0,\infty)} \sum_n |{dG_{n0}^0\over dt}(t)|$.
Then, for $t\leq T_1^+$,
\begin{eqnarray}\begin{array}{l}
\sum_n |u_n(t)-w_n(t)| \leq (S_1 +S_2) \varepsilon + 2 \tau  
|G_{n0}^0(T_0)|,   \\
\sum_n |u_n'(t)-w_n'(t) | \leq (S_3 +S_4) \varepsilon + 2 \tau  
|{dG_{n0}^0 \over dt}(T_0)|. \end{array} \label{tw10} 
\end{eqnarray}
The distances $|u_n(t)-w_n(t)|$, $|u_n'(t)-w_n'(t)|$
remain of order $\varepsilon$.
In particular, the oscillatory tail of $u_n(t)$ for $n>n_0$ is 
contained in the same band that contains $w_n(t)$ for $n>n_0$.

{\it Step 2: Generic stage.}
We iterate  Step 1 starting at times $T_{l}^+$, $l=1,2,...,$
according to the following induction procedure.
For a fixed $T_l^+$, (\ref{tw7}) holds for $n<n_0-l$, $n<n_1-l$,
$T_{l-1}^+\leq t \leq T_{l}^+$ and:
\begin{eqnarray}\begin{array}{l}
\sum_n |u_n(t)-w_n(t) |  \leq (S_1 +S_2) \varepsilon + 2 \tau S,  
 \\
\sum_n |u_n'(t)-w_n'(t) |  \leq (S_3 +S_4) \varepsilon + 2 \tau S
\end{array} \label{tw10bis}
\end{eqnarray}
holds  for  $t \leq T_{l}^+$, with $S={\rm Max} \big( \sum_{k} 
|G_{nk}^0(T_k)|,$ $  \sum_{k}|{dG_{nk}^0 \over dt}(T_k)| \big).$
Now, we shall show that these properties also hold for $T_{l+1}^+$.
 
For $T_l^+ \leq t\leq T_{l+1}^+$, the evolution of $w_n(t\pm \tau)$ 
and $u_n(t)$ is given by:
\begin{eqnarray}\begin{array}{l}
w_n(t\pm \tau)=\sum_k G_{nk}^0(t-T_{l}^+) w_k'
(T_{l}^+\pm \tau) +  \\
\sum_kG_{nk}^1(t-T_{l}^+) w_k(T_{l}^+\pm \tau)  + \\ 
F \int_{T_{l-1}^-}^{t} \sum_{k} 
G_{nk}^0(t  -z)  dz +   \\
\int_{T_{l}^+}^{t} \sum_{k> -l} 
G_{nk}^0(t  -z)  dz   +\\ 
 \sum_{k\leq -l}\int_{T_k^{\pm}}^{{\rm Max }
(t,T_k^{\pm})} G_{nk}^0(t  -z)  dz 
\end{array}\label{tw11}
\end{eqnarray}
\begin{eqnarray}\begin{array}{l}
u_n(t)=\sum_k G_{nk}^0(t-T_{l}^+) u_k'(T_{l}^+) +   \\
\sum_k G_{nk}^1(t-T_{l}^+) u_k(T_{l}^+)  + \\ 
F \int_{T_{l}^+}^{t}\!\! \sum_{k} 
G_{nk}^0(t  -z)  dz +  \\
\!\!\int_{T_{l}^+}^{t}\! \sum_{k> -l} 
G_{nk}^0(t  -z)  dz    +\\ 
 \sum_{k\leq -l}\int_{T_{l}^+}^{t} 
G_{nk}^0(t  -z)  H(u_{k}(z)-{1\over 2}) dz.
\end{array}\label{tw12}
\end{eqnarray}
Notice that we have taken as initial data the values of
$w_n(t\pm \tau)$ and $u_n(t)$ at time $T_l^+$. In this
way, formulae (\ref{tw11}) and (\ref{tw12}) only
involve the values of the Green functions in a short time 
interval $[0, {1\over |c|}+2\tau]$.
Since  ${1\over |c|}+2\tau>T^*$, the Green functions are 
both positive.
Recall that for this short time interval $G_{nk}^1$ and $G_{nk}^0$, 
together with their derivatives, take on large values for 
$k$ close to $n$. 
We can then use (\ref{tw7}) for $n<n_0-l, n<n_1-l$ at time
$T_l^+$, (\ref{tw10bis}) and (\ref{tw11})-(\ref{tw12}) to obtain
(\ref{tw7}) for $n<n_0-(l+1), n<n_1-(l+1)$ and
$T_{l}^+\leq t \leq T_{l+1}^+$. This means that $u_{-l}(t)-{1\over 2}$
changes sign at a time $t_l$ such that $T_l^+<t_l<T_l^- < T_{l+1}^+$.
We then obtain formula (\ref{tw6}) for $u_n(t)$ restricted to $t\leq
T_{l+1}^+$. Subtracting (\ref{tw1}) from (\ref{tw6}) for 
$t\leq T_{l+1}^+$, we find:
\begin{eqnarray}\begin{array}{l}
\sum_n |u_n-w_n |(t) \leq (S_1 +S_2) \varepsilon + 2 \tau  
\sum_{k\leq l} |G_{nk}^0(T_k)|  \\
\sum_n |u_n'-w_n' |(t) \leq (S_3 +S_4) \varepsilon + 2 \tau  
\sum_{k\leq l}|{dG_{nk}^0 \over dt}(T_k)|  
\end{array} \label{tw13}
\end{eqnarray}
This implies (\ref{tw10bis}) for $t\leq T_{l+1}^+$.
We are now ready to repeat the process starting at time
$T_{l+1}^+$.

{\it Step 3: Conclusion.} From Step 2
we obtain a sequence of times $t_l$ for 
$l=1,2,...$, with $T_{l}^+ < t_l <T_{l}^-$, at which
$u_{-l}(t)-{1\over 2}$ changes sign. 
In this way, we keep track of the times $t_l$ at which changes 
of sign take place and obtain formula (\ref{tw6}) for $u_n(t)$
for all $t$.
Subtracting (\ref{tw1}) from (\ref{tw6}) we find the
bound (\ref{tw07}) on $|u_n-w_n|$.

%\end{multi}
\end{document}